\documentclass[smallextended]{svjour3}
\RequirePackage{fix-cm}
\smartqed
\usepackage{graphicx}

\usepackage{amsmath,amsfonts}
\usepackage{algorithmic}
\usepackage{array}
\usepackage[caption=false,font=normalsize,labelfont=sf,textfont=sf]{subfig}
\usepackage{textcomp}
\usepackage{url}
\usepackage{graphicx}
\usepackage{adjustbox}
\usepackage{placeins}
\usepackage{subcaption}
\usepackage[noadjust]{cite}
\usepackage{float}
\usepackage{amsmath}
\usepackage{tipa}
\pdfglyphtounicode{tfm:tipa10/at}{0259}

\usepackage{balance}

\begin{document}

\title{A Terahertz Bandpass Filter Using a Capacitive Transition Circuit \\ and a Spoof Surface Plasmon Polariton Waveguide}

\author{Mohsen Haghighat$^{1,2}$ \and
	Levi Smith$^{1,2,*}$
    }

\institute{1. Department of Electrical and Computer Engineering, University of Victoria, Victoria, BC V8P 5C2, Canada.
\\
2. Centre for Advanced Materials and Related Technology (CAMTEC), University of Victoria, 3800 Finnerty Rd, Victoria, BC V8P 5C2, Canada.
\\
* Corresponding author;	\email{levismith@uvic.ca}
}

\date{Received: date / Accepted: date}

\maketitle

\begin{abstract}
This paper presents a novel terahertz (THz) bandpass filter (BPF) based on a spoof surface plasmon polariton (SSPP) waveguide with a center frequency of 1 THz and a \mbox{3 dB} bandwidth of 0.3 THz. The proposed BPF comprises cascaded high-pass and low-pass elements. The high-pass element is a capacitive gap in the SSPP transition circuit, and the low-pass element is the SSPP waveguide itself. We find that the measurement results, including cut-off frequencies, align well with the theoretical predictions and simulations. To the authors' knowledge, the proposed SSPP BPF is the first of its kind.

\end{abstract}

\section{Introduction}
Spoof surface plasmon polaritons (SSPP) devices operating at microwave and terahertz (THz) frequencies have attracted notable research attention \cite{Pendry2004_Mimicking,garcia-vidal_surfaces_2005,Review_of_SSPP_Tang_2019}, mainly due to their inheritable capabilities from typical surface plasmon polaritons (SPP), such as significant surface field confinement and dispersion properties which can be customized through geometric modifications  \cite{maier2007-plasmonics-book, huidobro_pendry_vidal_book_2018}.  SSPP structures were initially introduced using 3D geometries like an array of holes or grooves in a piece of metal that interacted with incident oblique light or waves \cite{Pendry2004_Mimicking,garcia-vidal_surfaces_2005}. Since then, the use of guided waves for SSPP excitation has been pursued due to their improved integration capabilities  \cite{Review_of_SSPP_Tang_2019}. A common guided-wave SSPP waveguide consists of a corrugated single-conductor that requires a matching or transition circuit. The majority of the literature focuses on the integration of SSPP in coplanar waveguides (CPW) \cite{ma2014broadband_10,unutmaz_terahertz_2019,unutmaz2020fixed_9,unutmaz_investigation_2022,mazdouri_miniaturized_2021,tang_concept_2019} due to compatibility with standard vector network analyzer probes. This involves connecting the CPW signal line to the SSPP and flaring the ground lines as the SSPP stub length increases gradually \cite{unutmaz_investigation_2022}. This configuration requires large areas of flaring ground planes in the transition circuit (TC), which consumes the area of the chip \cite{unutmaz_investigation_2022}.  

An SSPP waveguide can consist of more than one conductor; an example is the CPS-SSPP waveguide, where corrugations are introduced in each conductor of a standard CPS \cite{xu_terahertz_2019, xu_spoof_2020, Haghighat2024_NSRep}. An important benefit of the CPS-SSPP is the ease of integration with CPS feedlines that can be excited by photoconductive switching, which allows the investigation of devices at THz frequencies \cite{}. Like single-conductor SSPP waveguides, the transmission characteristics of a typical CPS-SSPP waveguide exhibit low-pass behavior that can be adjusted by the corrugation geometry. In this work, this low-pass behavior is used to control the upper-cut frequency of the BPF. Next, exciting the SSPP mode requires a TC. For the CPS-SSPP the TC consists of a gradual ramp in the corrugation depth. This work has a unique TC that interconnects a CPS feedline with a single-conductor SSPP waveguide. The combination of the TC and the SSPP waveguide forms the proposed BPF. 

The majority of the SSPP literature is focused on SSPP waveguiding with low loss throughout a limited bandwidth, such as \cite{unutmaz_terahertz_2019} or showing its low-pass filtering due to slow-wave properties such as \cite{guo_spoof_2018}. SSPP-based band-pass filters (BPF) have been reported in limited numbers at microwave and mm-wave frequencies \cite{guo_novel_BPF_IEEE_access_2018,Wang2019_BPF_SSPP_10GHz_IEEEAccess,Wei2020_BPF_SSPP_IEEE_plasma_sci,Liu-Xu2022_BPF_SSPP_80GHz_TMTT,Feng-Xu2024_BPF_SSPP_30GHz}; however, these structures generally use microwave cavities, which are difficult to integrate on a chip. In addition, there is a lack of experimentally validated SSPP-based BPF research at THz frequencies, mainly due to the lack of equipment and waveguiding challenges, such as different sources of loss at these frequencies
\cite{THz_Communication_challanges_IEEE_TTST_2021, Review_of_SSPP_Tang_2019}.

\begin{figure*}
    \centering
    \includegraphics[width=1\linewidth]{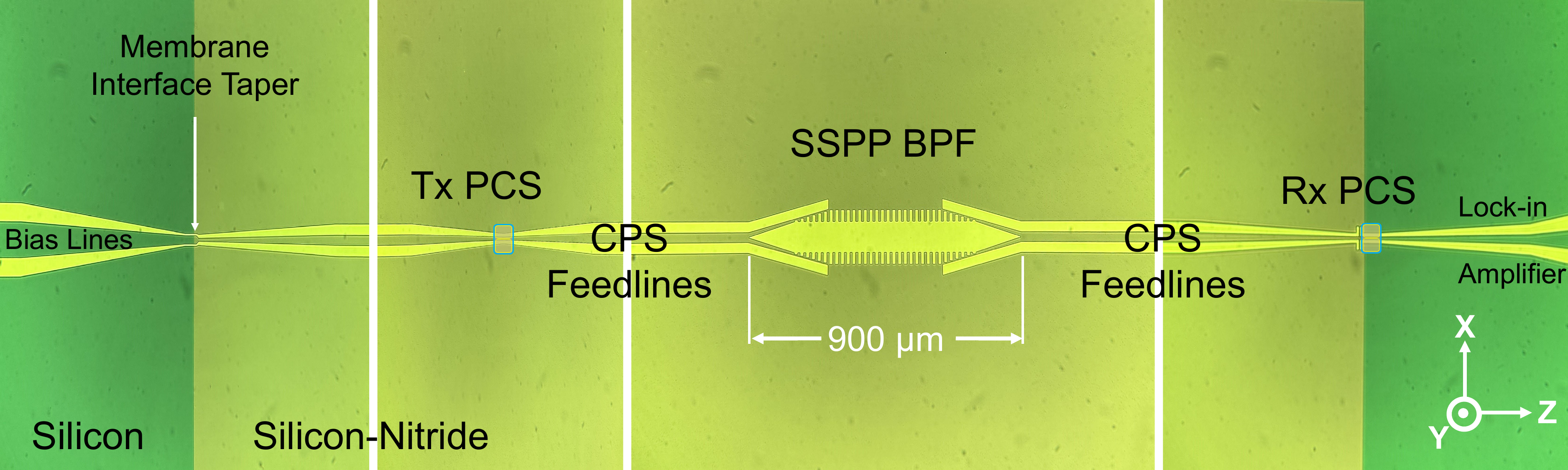}
    \caption{Fabricated THz SSPP BPF on thin Si-N membrane with CPS feedlines and transition circuits for excitation and Tx/Rx PCS for THz generation and detection.}
    \label{fig:BPF_V_Fabricated}
\end{figure*}

\begin{figure}
    \centering
    \includegraphics[width=0.6\linewidth]{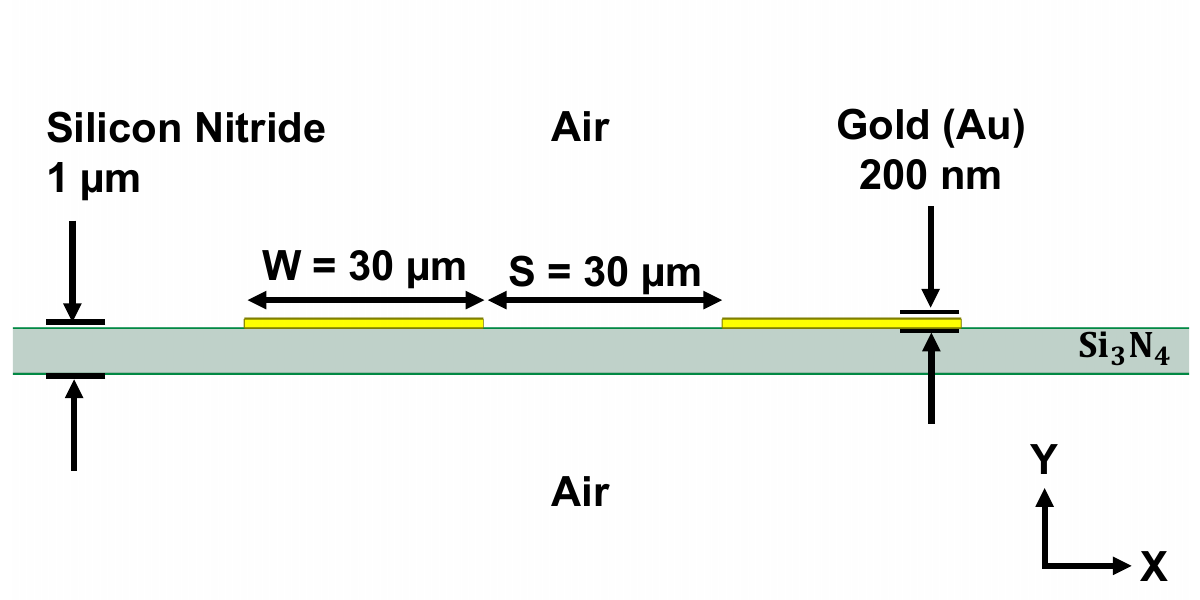}
    \caption{Cross section of CPS feedlines on the thin Silicon Nitride membrane.}
    \label{fig:BPF_Feedline_Cross_Section}
\end{figure}

This paper presents the experimental demonstration of a novel THz SSPP-based BPF structure 
by introducing the integration of CPS feedlines into a single conductor SSPP 
for the first time. It has been presented with simulation in our preliminary work at the 300 GHz band \cite{Haghighat_PACRIM_2024_6G_BPF}. For the integration of SSPP to CPS, a transition circuit (TC) is proposed to convert the quasi-TEM mode of the CPS to the SSPP TM mode \cite{unutmaz_investigation_2022, Haghighat2024_NSRep}. This is achieved by gradually extending the length of SSPP stubs while tapering the CPS lines and eventually terminating the CPS conductors. This TC configuration couples the propagating wave to the SSPP and blocks the low frequencies. To operate at THz frequencies, we employed a thin 1 \textmu m  silicon nitride (SiN) membrane, which mitigates loss and dispersion at THz frequencies \cite{Levi_Smith_CPS_on_Si3N4_1st,Haghighat2024_NSRep}. Figure \ref{fig:BPF_V_Fabricated} illustrates the proposed SSPP-based bandpass filter fabricated on the SiN membrane, together with photo-conductive switches (PCS) for THz generation and detection.  Figure \ref{fig:BPF_Feedline_Cross_Section} shows the cross-section of the CPS feedlines on the SiN membrane.

\section{Design and Simulation}
\label{sec:design}

ANSYS HFSS was used to assist in the design process and also to simulate the designed structure. Specifically, eigenmode simulations were used to obtain the dispersion curves for the SSPP waveguide unit cell. Frequency-domain (FD) simulations were also used to obtain the frequency-dependent S-parameters for the BPF. The simulated constitutive material parameters are as follows: $\varepsilon_{r,\text{SiN}}$ = 7.6, $\sigma_\text{SiN}=0$ and \mbox{tan $\delta_{\varepsilon,\text{SiN}}$ = 0.00526} \cite{Cataldo_Silicon_nitride_properties_2012}, and for the conductors $\sigma_{\text{Au}} = 4.1 \times 10^7$ S/m or $\sigma_{\text{cond}} = \infty$ depending on the simulation. For all simulations, the thickness of the SiN substrate is 1 \textmu m and the conductor thickness is 200 nm.

\begin{figure*}
    \centering       \includegraphics[width=1\linewidth]{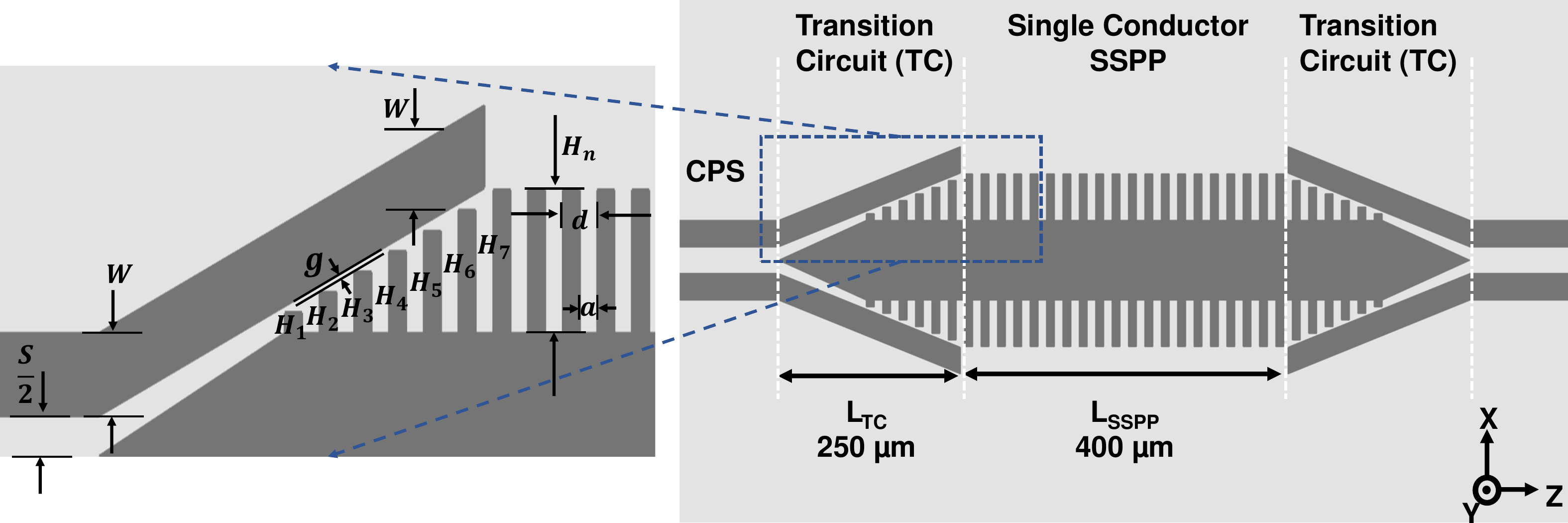}
    \caption{CPS to SSPP transition circuit dimensions for the proposed BPF }
    \label{BPF_V_TC}
\end{figure*}

The BPF presented in this work is a proof of concept selected to operate at THz frequencies. For this purpose, we select a center frequency of 1 THz and a bandwidth of 0.3 THz. For the design procedure, the BPF can be divided into cascaded low-pass and high-pass elements that are mostly independent of each other. The BPF consists of a single-conductor SSPP waveguide with a tapered CPS transition circuit (TC) as shown in Figure \ref{BPF_V_TC}. The TC acts as a high-pass filter and also performs mode conversion between the CPS TEM mode and the SSPP TM mode. The high-pass behavior is achieved using the capacitive gap between the CPS and the single-conductor SSPP waveguide. Mode conversion is achieved by gradually increasing the length of single-conductor SSPP stubs within the TC \cite{Haghighat2024_NSRep,ma2014broadband_10}. The lengths of the stubs are linearly tapered and tabulated in Table \ref{tab:TC_stubs}, where the row in bold font is a fabricated structure used in the experiment.

\begin{table}
\renewcommand{\arraystretch}{1.3}
\caption{Stub lengths for the TC (Units: \textmu m)}
\label{tab:TC_stubs}
\centering
\begin{tabular}{c | c c c c c c}
\hline
$H_n$ & $H_1$ & $H_2$ & $H_3$ & $H_4$ & $H_5$ & $H_6$  \\
\hline
35 & 5 & 10 & 15 & 20 & 25 & 30 \\

\textbf{42} & \textbf{6} & \textbf{12} & \textbf{18} & \textbf{24} & \textbf{30} & \textbf{36} \\

49 & 7 & 14 & 21 & 28 & 35 & 42 \\
\hline
\end{tabular}
\end{table}

\begin{figure}
    \centering
    \includegraphics[width=0.7\linewidth]{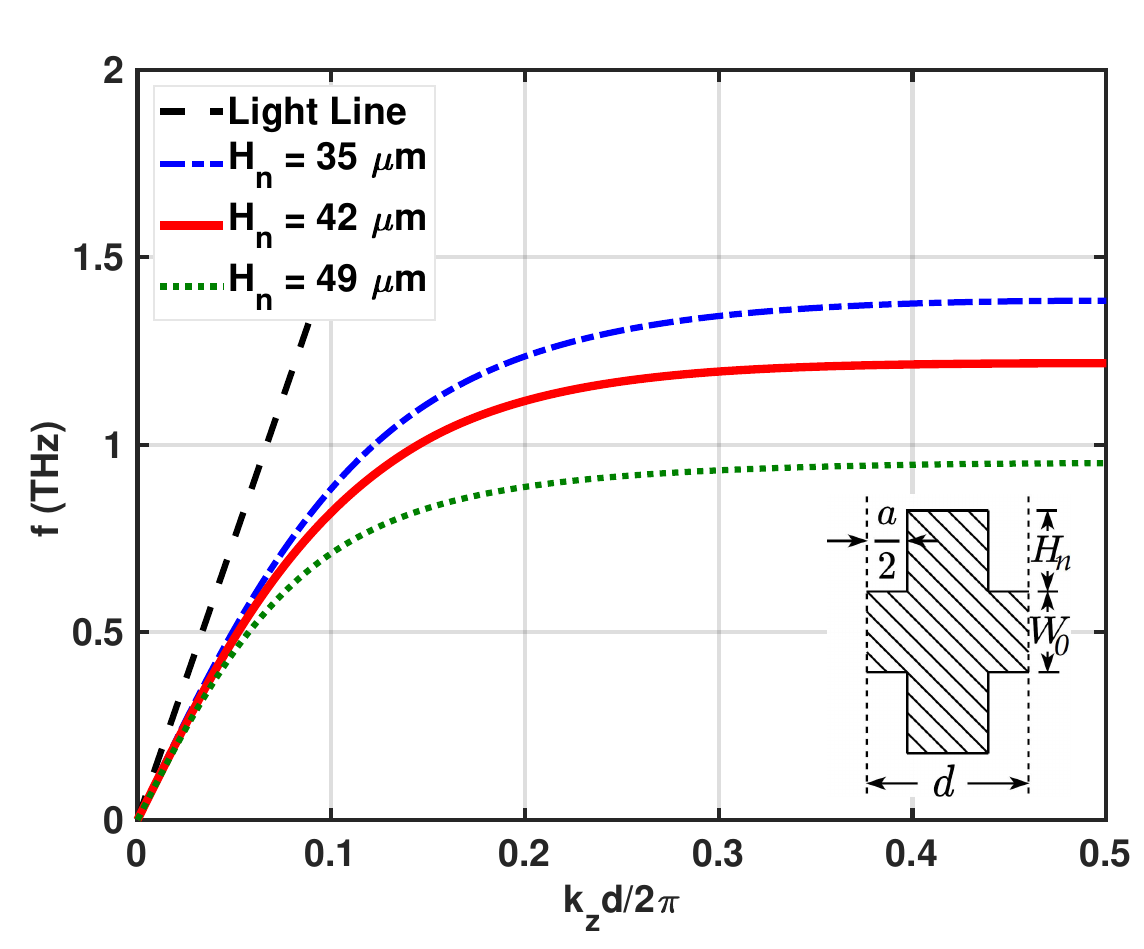}
    \caption{BPF Unit Cell Dispersion Curves, determining upper cut-off frequency. Unit cell dimensions: $d$ = 20 \textmu m, $a$ = 10 \textmu m, $W_0$ = 90 \textmu m, 21 \textmu m $\leq H_n \leq$ 70 \textmu m.}
    \label{fig:BPF_V_Dispersion}
\end{figure}

The upper cut-off frequency is controlled by the dimensions of the SSPP waveguide, in particular the heights of the stubs, $H_n$. We used an eigenmode simulation on the SSPP unit cell to extract the dispersion curves and the band-edge frequencies. Figure \ref{fig:BPF_V_Dispersion} plots the results of an eigenmode simulation for the unit cell shown in the inset. In this simulation, the period $d$ = 20 \textmu m, the aperture $a$ = 10 \textmu m, and the width of the central conductor, $W_0$ = 90 \textmu m, were fixed, and $H_n$ was varied to obtain the different curves. We selected a subwavelength value for $d$ and $a$ as they must be much smaller than the wavelength \cite{garcia-vidal_surfaces_2005,maier_terahertz_2006,shen_conformal_2013}. From Fig. \ref{fig:BPF_V_Dispersion}, it is clear that the band-edge frequency associated with $k_z d = \pi$ depends on $H_n$. In short, small values of $H_n$ correspond to higher values for the upper cut-off frequency. We note that the solid (red) trace in Fig. \ref{fig:BPF_V_Dispersion} corresponds to the fabricated structure.

The upper cut-off frequency can also be calculated by an analytical dispersion expression assuming that the SSPP shown in Fig. \ref{BPF_V_TC}
is modeled like a symmetric \mbox{($y$-$z$ plane)} 1D array of grooves with finite thickness by \cite{garcia-vidal_surfaces_2005, shen_conformal_2013}: 

\begin{equation}
k_z= k_{eff} \sqrt{1 + (a/d)^2\operatorname{tan}^2(k_{eff}H_n)},
\label{eqn:disp}
\end{equation}

\noindent where $k_{eff} =\omega \sqrt{\varepsilon_{eff}}/c$ is the effective wavenumber, $c$ is the speed of light, and $\varepsilon_{eff}$ is the effective relative permittivity which depends on $H_n$. From simulations, an SSPP waveguide on a thin SiN membrane, the effective permittivity can be approximated by $\varepsilon_{eff} =$ 2.15, 2.0, 1.85 when $H_n$ = 35, 42, 49 \textmu m, respectively.  Similarly to Fig. \ref{fig:BPF_V_Dispersion}, the solution to (\ref{eqn:disp}) shows that the band edge (and the associated upper cutoff frequency) of the SSPP structures are primarily controlled by $H_n$, thus (\ref{eqn:disp}) provides a useful starting point and theoretical aid in the design of SSPP filters \cite{ma2014broadband_10, guo_spoof_2018}. However, we recommend that numerical eigenmode simulations be used prior to actual device fabrication, as they provide a more accurate prediction of device performance. Lastly, we note that the solution to (\ref{eqn:disp}) for $H_n$ = 42 \textmu m predicts an upper cut-off frequency of 1.2 THz.

We used a 1 \textmu m thick SiN substrate, which has been selected to minimize loss and dispersion \cite{Levi_Smith_CPS_on_Si3N4_1st}; however, other substrates can be used to implement this structure, expecting higher loss at THz frequencies. Note that using other substrates changes the effective permittivity and the band-edge locations; hence, the relation between the geometry and the band-edge should be recalculated.

The lower cut-off frequency is determined primarily by the gap between the tapered CPS feedlines and the single-conductor SSPP in the TC, which is indicated as $g$ in Fig. \ref{BPF_V_TC}. Reducing the gap between the tapered CPS and TC stubs increases the series capacitance, thereby reducing the lower cut-off frequency. In contrast, increasing the gap decreases the capacitance, which increases the lower cut-off frequency. However, increasing the gap also reduces the coupling and increases the insertion loss. We selected $g =$ 4 \textmu m to have an acceptable passband insertion loss (IL), which was found to be $\approx$1.5 dB from simulation and a bandwidth of 0.3 THz. Another method of controlling the lower cut-off frequency is to increase the length of the TC coupling region. This is denoted as $L_{TC}$ in Fig. \ref{BPF_V_TC}. Larger values of $L_{TC}$ will enable wider bandwidths.

\begin{figure}
    \centering
    \includegraphics[width=0.7\linewidth]{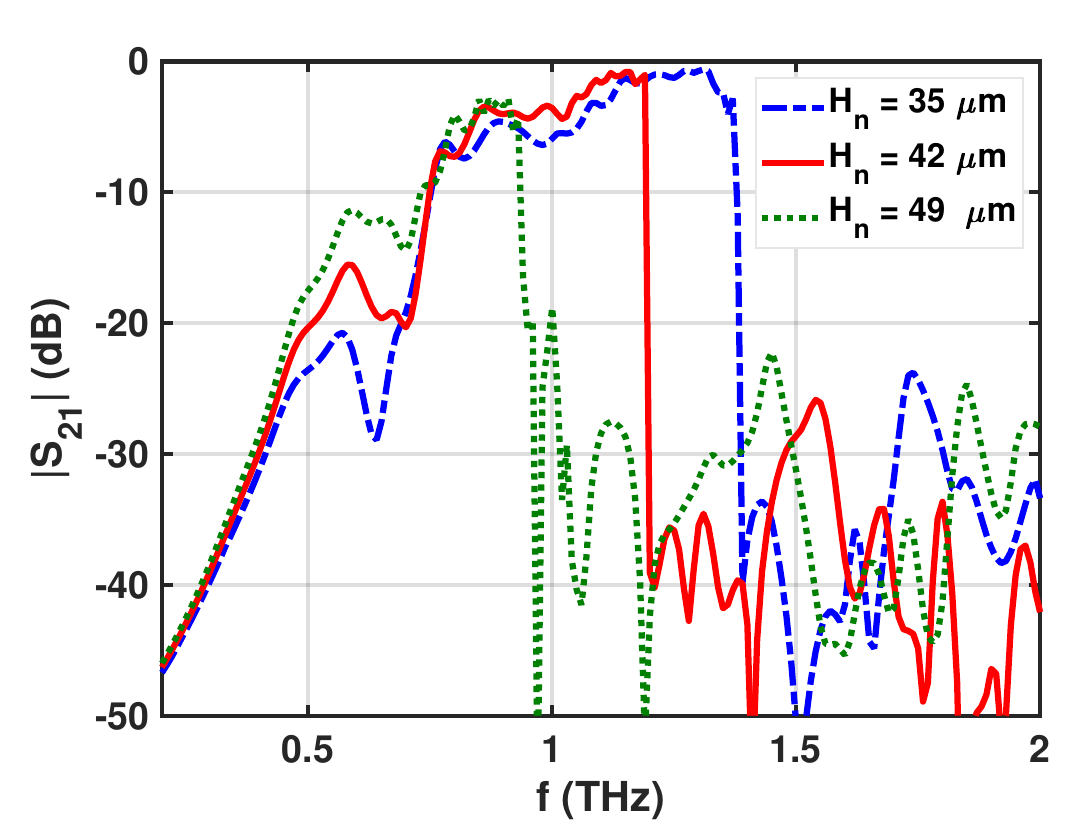}
    \caption{Transmission response ($S_{21}$) of the SSPP BPF structure  with variable $H_n$ from 28 \textmu m to 70  \textmu m when the conductor is modeled as PEC.}
    \label{fig:BPF_V_S21_Variable_H}
\end{figure}

As mentioned, the objective of the proposed filter is to combine low-pass and high-pass responses to produce a BPF by integrating the CPS feedlines, the TCs and the single-conductor SSPP waveguide, as shown in Fig. \ref{BPF_V_TC}. To predict the broadband response of the BPF numerical simulations should be used. First, we look at the transmission response for different values of $H_n$ when the conductors are modeled as PECs. The result is plotted in Fig. \ref{fig:BPF_V_S21_Variable_H}. For these simulations, $d = $ 20 \textmu m, $a$ = 10 \textmu m, $S$ = 30 \textmu m, and $W$ = 30 \textmu m. From Fig. \ref{fig:BPF_V_S21_Variable_H} it is clear that the upper cut-off frequency is dependent on the value of $H_n$ which is aligned with the results of the eigenmode simulation for the unit cell plotted in Fig. \ref{fig:BPF_V_Dispersion}.

\begin{figure}
    \centering
    \includegraphics[width=0.7\linewidth]{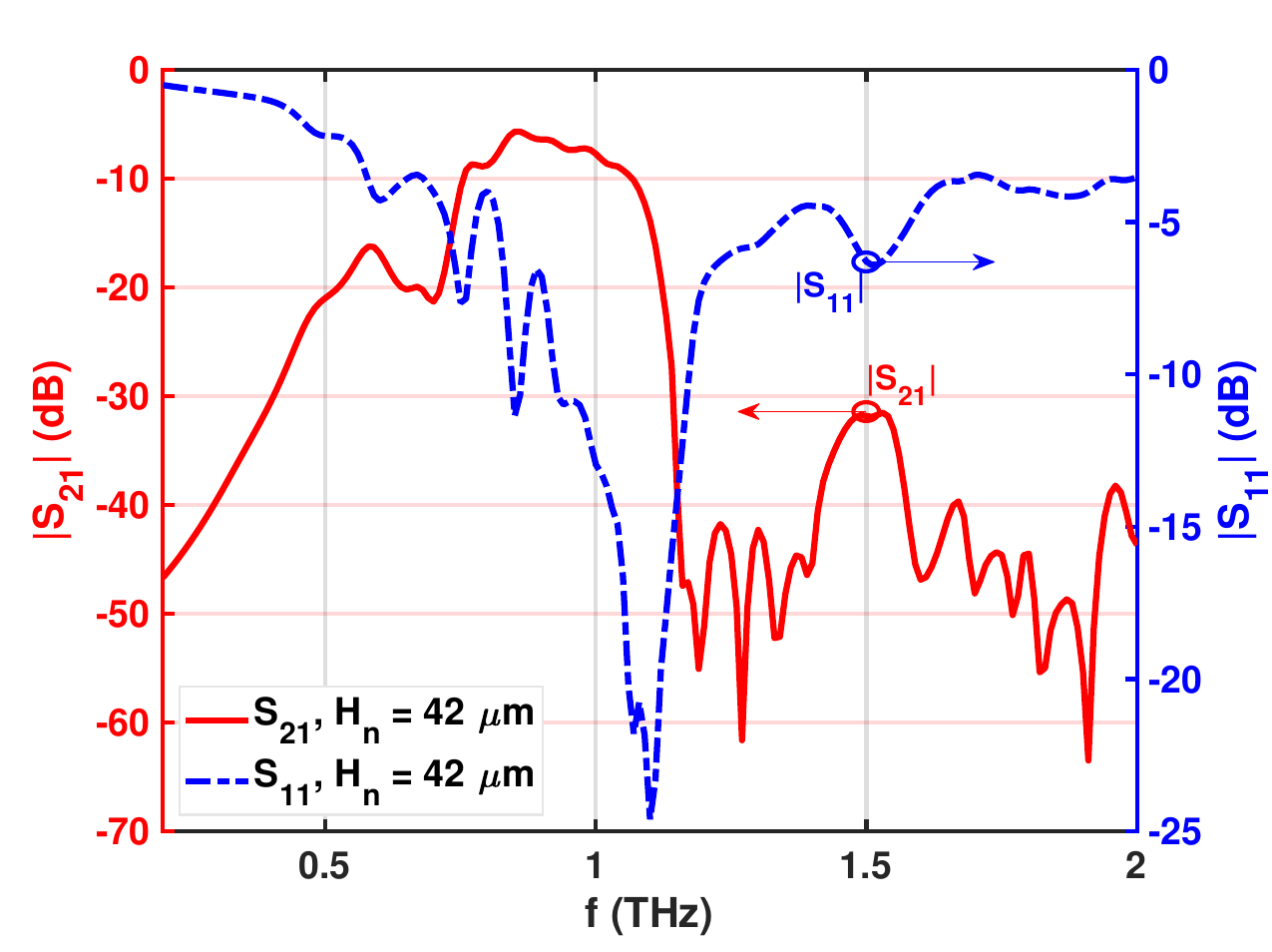}
    \caption{$S_{21}$ and $S_{11}$ of the BPF structure  with $H_n$ = 42 \textmu m considering dielectric and conductor losses.}
    \label{fig:BPF_V_S21_S11_H42um}
\end{figure}

Next, we model the 200 nm thick conductors as gold to observe the effect of a finite conductivity. The S-parameters for the BPF are shown in Fig. \ref{fig:BPF_V_S21_S11_H42um} which illustrates a passband insertion loss of 5-7 dB.

\begin{figure}
    \centering
    \includegraphics[width=0.8\linewidth]{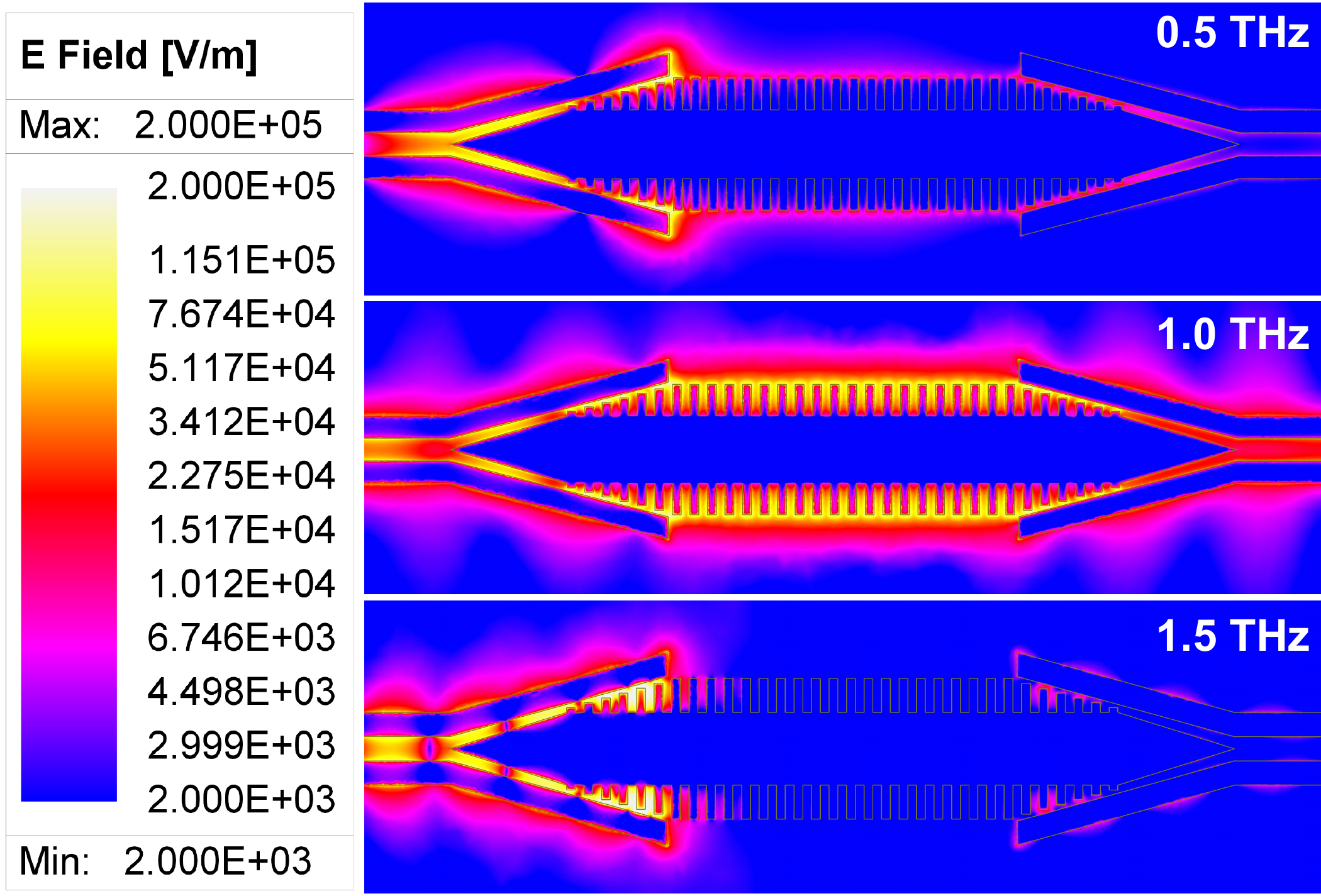}
    \caption{Field plots of the proposed SSPP BPF structure with $H_n = 42$ \textmu m at 0.5 THz (lower stopband), 1 THz (passband), and 1.5 THz (upper stopband). The temperature (color) scale ranges from blue ($2 \times 10^3$ V/m) to white ($2 \times 10^5$ V/m) on a logarithmic scale.}
    \label{fig:BPF_V_THz_Field_Plots}
\end{figure}

For visualization of filter operation, Fig. \ref{fig:BPF_V_THz_Field_Plots}
illustrates the electric field plots for the structure with $H_n$ = 42 \textmu m at 1 THz in the passband and lower/upper stopbands at 0.5 and 1.5 THz. The field plot at 1 THz also illustrates the field confinement capability of the SSPP structure in the groove areas.

\section{Experiment}
\begin{figure}
    \centering
    \includegraphics[width=0.7\linewidth]{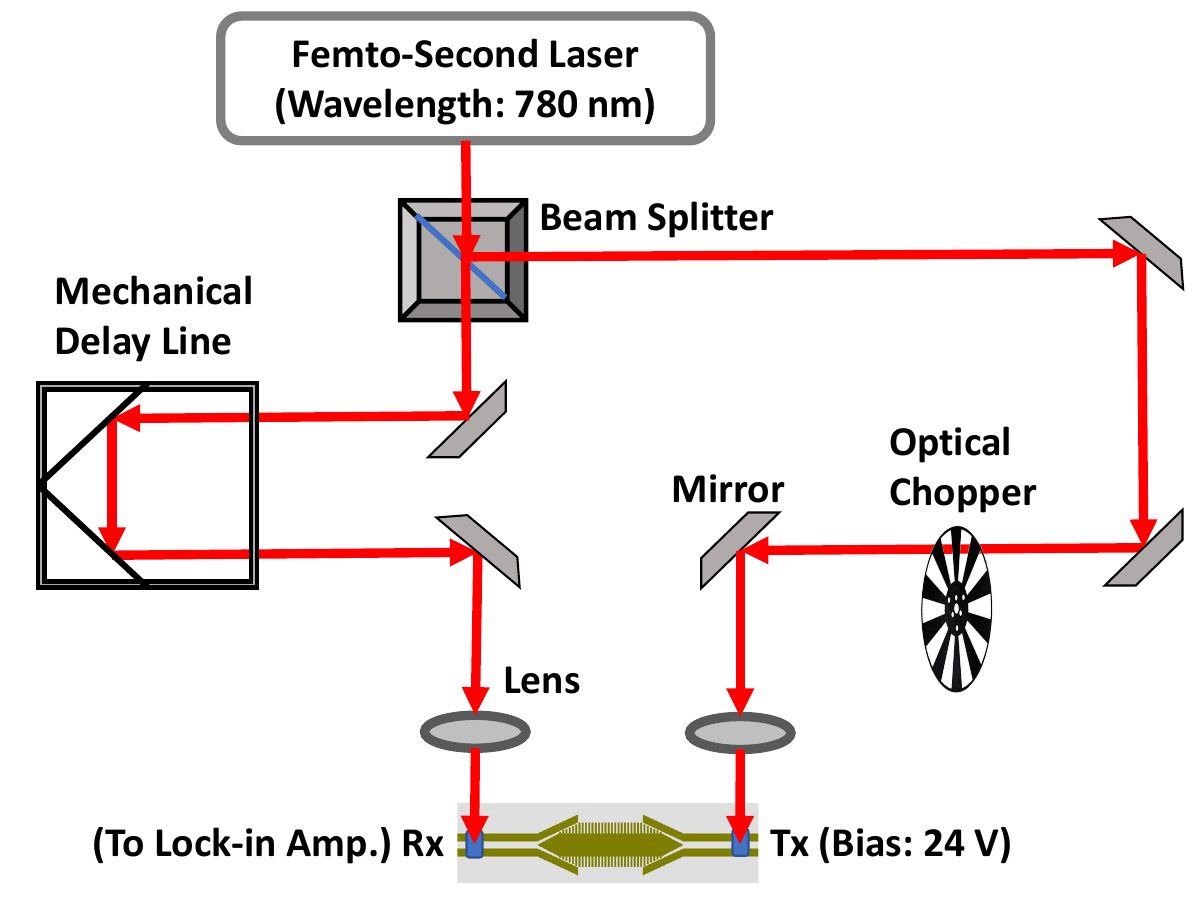}
    \caption{BPF measurement setup based on modified THz time domain spectroscopy}
    \label{fig:BPF_V_Meas_Setup_Diagram}
\end{figure}

The experimental setup for the measurements is a modified THz time-domain spectroscopy (THz-TDS) setup presented in \cite{Levi_Smith_CPS_on_Si3N4_1st}, as depicted in Fig. \ref{fig:BPF_V_Meas_Setup_Diagram}. This setup utilizes a femtosecond pulsed laser with specific parameters: a wavelength of 780 nm, a pulse width of 90 fs, a pulse rate of 80 MHz, and an average output power of 27 mW. The laser beam is focused onto the photoconductive switches (PCS) placed on the designated transmitter (Tx) and receiver (Rx) spots on the feedlines (see Fig. \ref{fig:BPF_V_Fabricated}), serving as transmitter and receiver, to generate and detect THz pulses with photoconductive switching and sampling methods \cite{Levi_Smith_CPS_on_Si3N4_1st}.

The PCS, with dimensions of 70 \textmu m $\times$ 40 \textmu m $\times$ 1.5 \textmu m and a 5 \textmu m gap between the metal contacts of the PCS (gold), are bonded using water droplets with Van der Waals forces (VDW), as explained in \cite{VDW1990}. The transmitted signal is then reconstructed by adjusting a mechanical delay line and measuring the receiver current using a lock-in amplifier, following techniques similar to those employed in conventional THz-TDS setups \cite{Levi_Smith_CPS_on_Si3N4_1st,gomaa_terahertz_2020}.

The fabrication of photo-conductive switches (PCSs) involves a multi-step process. Initially, a low-temperature grown gallium arsenide (LT-GaAs) layer is deposited on a sacrificial aluminum arsenide (AlAs) layer, which is situated on a semi-insulating GaAs substrate. Subsequently, the LT-GaAs surface undergoes photolithography to establish Au contacts. Each PCS region is then masked and subjected to wet etching using citric acid and hydrogen peroxide to define the PCS thickness. After cleaning and re-masking with etch-resist wax, the LT-GaAs layer is dissolved in hydrofluoric acid (HF), separating it from the AlAs layer. Additional steps are taken to disconnect the remaining LT-GaAs film, resulting in the formation of multiple active regions of LT-GaAs PCSs \cite{Rios2015_bowtie_PCA}.

\section{Measurement Results and Discussion}

\begin{figure*}
    \centering
       \includegraphics[width=1\linewidth]{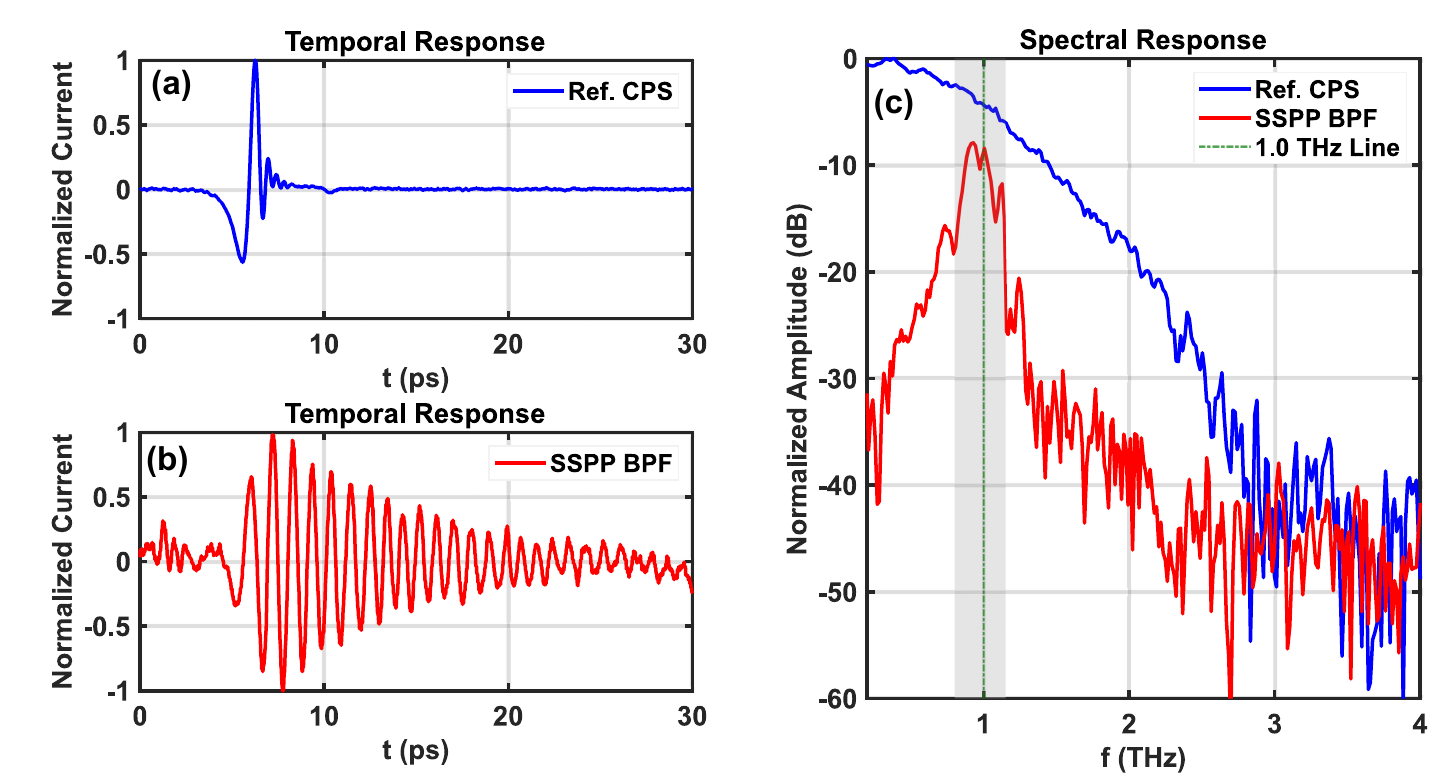}
    \caption{Measurement results for the proposed SSPP-based BPF and corresponding reference CPS}
    \label{BPF_V_Measurement}
\end{figure*}

The experimental results of a reference CPS waveguide and SSPP-based BPF with $H_n$ = 42 \textmu m are shown in Fig. \ref{BPF_V_Measurement}(a,b). The reference CPS is a feedline shown in Fig. \ref{fig:BPF_V_Fabricated} of the same length. The temporal response of CPS is a transient subpicosecond pulse (Fig. \ref{BPF_V_Measurement}a) that will have a broad spectral content. For the BPF, the temporal response is oscillatory (Fig. \ref{BPF_V_Measurement}b), aligned with a narrow-band signal. As expected, the oscillation period is $\approx$ 1 ps which corresponds to 1 THz, which is the center frequency of the BPF device under test. Figure  \ref{BPF_V_Measurement}c displays the corresponding spectral responses obtained by applying the Discrete Fourier Transform (DFT) to the temporal responses.  The spectral response of the BPF shows sharp reductions at 0.8 and 1.2 THz, which are aligned with the simulated predictions shown in Fig. \ref{fig:BPF_V_S21_S11_H42um}.

The comparison of spectral responses in Fig. \ref{BPF_V_Measurement} shows 15-20 dB out-of-band rejection for BPF, which is lower than expected, which should be more than 20 dB for low frequencies and more than 30 dB for upper band rejection. This is primarily because of noise levels and the weakness of the BPF output since it is a relatively narrow-band signal and holds only a small portion of the total pulse energy. Thus, the low signal-to-noise ratio disrupts the resolution of the real out-of-band rejection of the filter. Despite the weakness of the signal power, the measured out-of-band rejection is acceptable. Another important parameter to be extracted from the measurement results is the difference between the amplitude of the spectrum between the BPF and CPS reference in the passband, which is $\approx\ $5 dB that aligns with the minimum insertion loss expectations from the S-parameter simulations (see Fig. \ref{fig:BPF_V_S21_S11_H42um}). 

\begin{figure}
    \centering
    \includegraphics[width=0.6\linewidth]{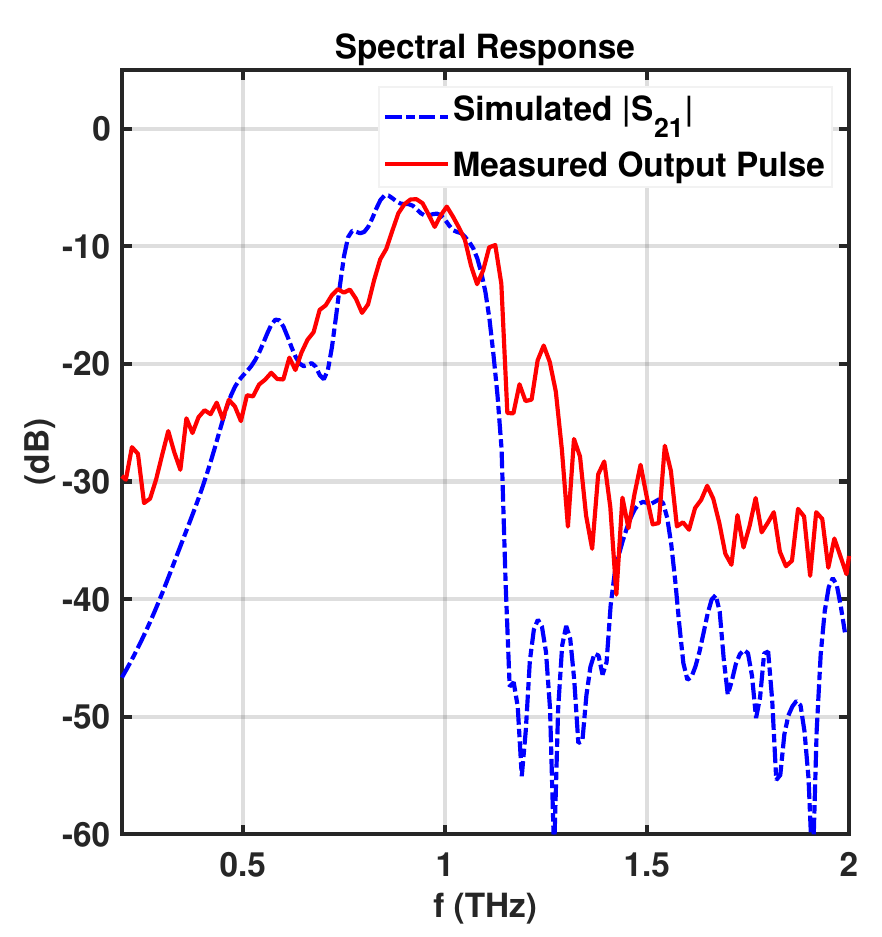}
    \caption{Simulated   $|S_{21}|$ vs normalized amplitude output measured pulse (in dB) for the structure with $H_n$ = 42 \textmu m.}
    \label{fig:BPF_V_Meas_vs_S21}
\end{figure}

We also compared the simulated $S_{21}$ with the measured pulse normalized to the simulated passband transmission to illustrate the operation of the device in Fig. \ref{fig:BPF_V_Meas_vs_S21}. Comparison between simulated and experimental spectral responses shows reasonable agreement and demonstrates that our proof-of-concept filter is a promising candidate for future THz BPFs.

It is worth mentioning that the spectral responses have inherent amplitude decay because these are Fourier transforms of finite-duration transient pulses rather than impulses with a flat spectrum \cite{Levi_Smith_CPS_on_Si3N4_1st,Haghighat2024_NSRep}. Therefore, the spectrum of BPF has a baseline that resembles the reference spectral curve. The fluctuations in the measurement are attributed to uncorrelated system noise sources, including those from the laser, equipment, and environment, as well as fluctuations in the PCS photo-current due to fabrication imperfections. These effects were also observed in the measurement of the CPS reference, where we expect a much smoother spectral response based on simulations.

\section{Conclusion}
This paper introduces a novel terahertz (THz) spoof surface plasmon polariton (SSPP) bandpass filter, marking a significant milestone in SSPP-based filter structures by integrating a single-conductor SSPP to CPS. Through experimental validation, the filter demonstrates impressive performance metrics, including a low passband amplitude difference of 5 dB compared to a CPS feedline with the same length and dimensions, and out-of-band rejection surpassing 20 dB for low frequencies and about 15-20 dB for high frequencies. With a 3 dB bandwidth of approximately 0.3 THz, ranging from 0.87 to 1.17 THz, the filter exhibits robust signal transmission and rejection capabilities within its passband and stopbands, closely aligned with simulations. Notably, the integration of a single-conductor SSPP with a dual-conductor CPS feedline represents a pioneering approach. The guided-wave transmission of the THz signal was enabled by the use of a thin 1 \textmu m Silicon Nitride membrane to mitigate loss and dispersion within the THz range. This study holds promise for a wide range of applications in next-generation communication systems, interconnects, and material sensing, highlighting the significance of planar SSPP structures in advancing THz technologies.

\section*{Acknowledgments}

This work was supported by an NSERC Discovery Grant. The authors thank 4D LABS at Simon Fraser University for the fabrication of the CPS waveguides and the thin membrane, and also the Center for Advanced Materials and Related Technology (CAMTEC) at the University of Victoria for providing Nanofab facilities for the fabrication of the PCS devices.

\vspace{5pt}

\noindent \textbf{Funding} This work was supported by the Natural Sciences and Engineering Research Council of Canada (NSERC) (RGPIN-2022-03277).

\noindent \textbf{Availability of data and materials} Data supporting the findings of this study are available from the corresponding author L.S. on reasonable request.

\section*{Declarations}

\noindent \textbf{Ethical Approval} Not applicable.

\vspace{8pt}

\noindent \textbf{Competing Interests} The authors declare no competing interests.

\bibliographystyle{unsrt}
\bibliography{SSPP_references}

@article{Cataldo_Silicon_nitride_properties_2012,
author = {Giuseppe Cataldo and James A. Beall and Hsiao-Mei Cho and Brendan McAndrew and Michael D. Niemack and Edward J. Wollack},
journal = {Optics Letters},
number = {20},
pages = {4200--4202},
publisher = {Optica Publishing Group},
title = {Infrared dielectric properties of low-stress silicon nitride},
volume = {37},
month = {Oct},
year = {2012},
uurl = {https://opg.optica.org/ol/abstract.cfm?URI=ol-37-20-4200},
doi = {10.1364/OL.37.004200},
}

@article{Review_of_SSPP_Tang_2019,
author = {Tang, Wen Xuan and Zhang, Hao Chi and Ma, Hui Feng and Jiang, Wei Xiang and Cui, Tie Jun},
title = {Concept, Theory, Design, and Applications of Spoof Surface Plasmon Polaritons at Microwave Frequencies},
journal = {Advanced Optical Materials},
volume = {7},
number = {1},
pages = {1800421},
doi = {https://doi.org/10.1002/adom.201800421},
uuurrrllll = {https://onlinelibrary.wiley.com/doi/abs/10.1002/adom.201800421},
eprint = {https://onlinelibrary.wiley.com/doi/pdf/10.1002/adom.201800421},
year = {2019}
}

@book{huidobro_pendry_vidal_book_2018, 
place={Cambridge}, 
series={Elements in Emerging Theories and Technologies in Metamaterials}, 
title={Spoof Surface Plasmon Metamaterials}, DOI={10.1017/9781108553445}, 
publisher={Cambridge University Press}, 
author={Huidobro, Paloma Arroyo and Fernández-Domínguez, Antonio I. and Pendry, John B. and Martín-Moreno, Luis and Garcia-Vidal, Francisco J.}, 
year={2018}, 
collection={Elements in Emerging Theories and Technologies in Metamaterials}
}

@article{xu_terahertz_2019,
	title = {Terahertz broadband spoof surface plasmon polaritons using high-order mode developed from ultra-compact split-ring grooves},
	volume = {27},
	issn = {1094-4087},
	uuurrrlll = {https://opg.optica.org/abstract.cfm?URI=oe-27-4-4354},
	doi = {10.1364/OE.27.004354},
	language = {en},
	number = {4},
	uuurrrllldate = {2023-01-15},
	journal = {Optics Express},
	author = {Xu, Kai-Da and Guo, Ying Jiang and Deng, Xianjin},
	month = feb,
	year = {2019},
	pages = {4354},
}

@article{xu_spoof_2020,
	title = {Spoof {Surface} {Plasmon} {Polaritons} {Based} on {Balanced} {Coplanar} {Stripline} {Waveguides}},
	volume = {32},
	issn = {1041-1135, 1941-0174},
	uuurrrlll = {https://ieeexplore.ieee.org/document/8918271/},
	doi = {10.1109/LPT.2019.2957059},
	language = {en},
	number = {1},
	uuurrrllldate = {2023-01-15},
	journal = {IEEE Photonics Technology Letters},
	author = {Xu, Kai-Da and Zhang, Fengyu and Guo, Yingjiang and Ye, Longfang and Liu, Yanhui},
	month = jan,
	year = {2020},
	pages = {55--58},
}

@article{guo_spoof_2018,
	title = {Spoof plasmonic waveguide developed from coplanar stripline for strongly confined terahertz propagation and its application in microwave filters},
	volume = {26},
	issn = {1094-4087},
	uuurrrlll = {https://opg.optica.org/abstract.cfm?URI=oe-26-8-10589},
	doi = {10.1364/OE.26.010589},
	language = {en},
	number = {8},
	uuurrrllldate = {2023-01-15},
	journal = {Optics Express},
	author = {Guo, Ying Jiang and Da Xu, Kai and Tang, Xiaohong},
	month = apr,
	year = {2018},
	pages = {10589},
}

@article{unutmaz_investigation_2022,
	title = {Investigation of the {Transitions} for {Coplanar} {Waveguide} to {Terahertz} {Spoof} {Surface} {Plasmon} {Polariton} {Waveguides}},
	volume = {70},
	issn = {0018-926X, 1558-2221},
	uuurrrlll = {https://ieeexplore.ieee.org/document/9615007/},
	doi = {10.1109/TAP.2021.3126369},
	language = {en},
	number = {4},
	uuurrrllldate = {2023-01-15},
	journal = {IEEE Transactions on Antennas and Propagation},
	author = {Unutmaz, Muhammed Abdullah and Ozsahin, Gulay and Abacilar, Tuna and Unlu, Mehmet},
	month = apr,
	year = {2022},
	pages = {3002--3010},
}

@article{Levi_Smith_CPS_on_Si3N4_1st,
author = {Levi Smith and Thomas Darcie},
journal = {Optics Express},
number = {10},
pages = {13653--13663},
publisher = {Optica Publishing Group},
title = {Demonstration of a low-distortion terahertz system-on-chip using a CPS waveguide on a thin membrane substrate},
volume = {27},
month = {May},
year = {2019},
uuurrrlll = {https://opg.optica.org/oe/abstract.cfm?URI=oe-27-10-13653},
doi = {10.1364/OE.27.013653},
}

@article{mazdouri_miniaturized_2021,
	title = {Miniaturized spoof {SPPs} filter based on multiple resonators or {5G} applications},
	volume = {11},
	issn = {2045-2322},
	uuurrrlll = {https://www.nature.com/articles/s41598-021-01944-6},
	doi = {10.1038/s41598-021-01944-6},
	language = {en},
	number = {1},
	uuurrrllldate = {2023-01-15},
	journal = {Scientific Reports},
	author = {Mazdouri, Behnam and Honari, Mohammad Mahdi and Mirzavand, Rashid},
	month = nov,
	year = {2021},
	pages = {22557},
}

@article{guo_novel_BPF_IEEE_access_2018,
	title = {Novel Surface Plasmon Polariton Waveguides with Enhanced Field Confinement for Microwave-Frequency Ultra-Wideband Bandpass Filters},
	volume = {6},
	issn = {2169-3536},
	uuurrrlll = {http://ieeexplore.ieee.org/document/8301436/},
	doi = {10.1109/ACCESS.2018.2808335},
	language = {en},
	uuurrrllldate = {2023-01-15},
	journal = {IEEE Access},
	author = {Guo, Ying Jiang and Xu, Kai Da and Liu, Yanhui and Tang, Xiaohong},
	year = {2018},
	pages = {10249--10256},
	
}

@article{tang_concept_2019,
	title = {Concept, {Theory}, {Design}, and {Applications} of {Spoof} {Surface} {Plasmon} {Polaritons} at {Microwave} {Frequencies}},
	volume = {7},
	issn = {21951071},
	uuurrrlll = {https://onlinelibrary.wiley.com/doi/10.1002/adom.201800421},
	doi = {10.1002/adom.201800421},
	language = {en},
	number = {1},
	uuurrrllldate = {2023-01-15},
	journal = {Advanced Optical Materials},
	author = {Tang, Wen Xuan and Zhang, Hao Chi and Ma, Hui Feng and Jiang, Wei Xiang and Cui, Tie Jun},
	month = jan,
	year = {2019},
	pages = {1800421},
}

@article{unutmaz_terahertz_2019,
	title = {Terahertz {Spoof} {Surface} {Plasmon} {Polariton} {Waveguides}: {A} {Comprehensive} {Model} with {Experimental} {Verification}},
	volume = {9},
	issn = {2045-2322},
	shorttitle = {Terahertz {Spoof} {Surface} {Plasmon} {Polariton} {Waveguides}},
	uuurrrlll = {https://www.nature.com/articles/s41598-019-44029-1},
	doi = {10.1038/s41598-019-44029-1},
	language = {en},
	number = {1},
	uuurrrllldate = {2023-01-15},
	journal = {Scientific Reports},
	author = {Unutmaz, Muhammed Abdullah and Unlu, Mehmet},
	month = may,
	year = {2019},
	pages = {7616},
}

@article{gomaa_terahertz_2020,
	title = {Terahertz low-pass filter based on cascaded resonators formed by {CPS} bending on a thin membrane},
	volume = {28},
	issn = {1094-4087},
	uuurrrlll = {https://opg.optica.org/abstract.cfm?URI=oe-28-21-31967},
	doi = {10.1364/OE.403702},
	language = {en},
	number = {21},
	uuurrrllldate = {2023-01-15},
	journal = {Optics Express},
	author = {Gomaa, Walid and Smith, Levi and Shiran, Vahid and Darcie, Thomas},
	month = oct,
	year = {2020},
	pages = {31967},
}

@article{maier_terahertz_2006,
	title = {Terahertz {Surface} {Plasmon}-{Polariton} {Propagation} and {Focusing} on {Periodically} {Corrugated} {Metal} {Wires}},
	volume = {97},
	issn = {0031-9007, 1079-7114},
	uuurrrlll = {https://link.aps.org/doi/10.1103/PhysRevLett.97.176805},
	doi = {10.1103/PhysRevLett.97.176805},
	language = {en},
	number = {17},
	uuurrrllldate = {2023-01-15},
	journal = {Physical Review Letters},
	author = {Maier, Stefan A. and Andrews, Steve R. and Martín-Moreno, L. and García-Vidal, F. J.},
	month = oct,
	year = {2006},
	pages = {176805},
}

@article{
Pendry2004_Mimicking,
author = {J. B. Pendry  and L. Martín-Moreno  and F. J. Garcia-Vidal },
title = {Mimicking Surface Plasmons with Structured Surfaces},
journal = {Science},
volume = {305},
number = {5685},
pages = {847-848},
year = {2004},
doi = {10.1126/science.1098999},
UURL = {https://www.science.org/doi/abs/10.1126/science.1098999},
eprint = {https://www.science.org/doi/pdf/10.1126/science.1098999}}

@article{unutmaz2020fixed_9,
title={Fixed physical length spoof surface plasmon polariton delay lines for a 2-bit phase shifter},
author={Unutmaz, M. A. and Unlu, M.},
journal={J. Opt. Soc. Am. B},
volume={37},
number={4},
pages={1116--1121},
year={2020},
month={Mar.}
}

@article{ma2014broadband_10,
title={Broadband and high-efficiency conversion from guided waves to spoof surface plasmon polaritons},
author={Ma, H. F. and Shen, X. and Cheng, Q. and Jiang, W. X. and Cui, T. J.},
journal={Laser Photonics Rev.},
volume={8},
number={1},
pages={146-151},
year={2014},
month={Jan.}
}

@book{maier2007-plasmonics-book,
  title={Plasmonics: fundamentals and applications},
  author={Maier, Stefan A and others},
  volume={1},
  year={2007},
  publisher={Springer}
}

@article{Rios2015_bowtie_PCA,
year = {2015},
volume = {17},
number = {12},
pages = {125802},
author = {Rubén Darío Velásquez Ríos and Siméon Bikorimana and Muhammad Ali Ummy and Roger Dorsinville and Sang-Woo Seo},
title = {A bow-tie photoconductive antenna using a low-temperature-grown GaAs thin-film on a silicon substrate for terahertz wave generation and detection},
journal = {Journal of Optics},
}

@article{VDW1990,
    author = {Yablonovitch, E. and Hwang, D. M. and Gmitter, T. J. and Florez, L. T. and Harbison, J. P.},
    title = "{Van der Waals bonding of GaAs epitaxial liftoff films onto arbitrary substrates}",
    journal = {Applied Physics Letters},
    volume = {56},
    number = {24},
    pages = {2419-2421},
    year = {1990},
    month = {06},
}

@article{shen_conformal_2013,
	title = {Conformal surface plasmons propagating on ultrathin and flexible films},
	volume = {110},
	number = {1},
	journal = {Proceedings of the National Academy of Sciences},
	author = {Shen, Xiaopeng and Cui, Tie Jun and Martin-Cano, Diego and Garcia-Vidal, Francisco J.},
	year = {2013},
	pages = {40--45},
}

@article{garcia-vidal_surfaces_2005,
	title = {Surfaces with holes in them: new plasmonic metamaterials},
	volume = {7},
	number = {2},
	journal = {Journal of Optics A: Pure and Applied Optics},
	author = {Garcia-Vidal, F. J. and Martín-Moreno, L. and Pendry, J. B.},
	year = {2005},
	pages = {S97},
}

@article{Haghighat2024_NSRep, author = {Haghighat, M. and Darcie, T. and Smith, L.}, title = {Demonstration of a terahertz coplanar-strip spoof-surface-plasmon-polariton low-pass filter}, journal = {Scientific Reports}, year = {2024}, volume = {14}, issue = {1}, doi = {10.1038/s41598-023-50599-y} }

@ARTICLE{Wang2019_BPF_SSPP_10GHz_IEEEAccess,
  author={Wang, Jun and Zhao, Lei and Hao, Zhang-Cheng},
  journal={IEEE Access}, 
  title={A Band-Pass Filter Based on the Spoof Surface Plasmon Polaritons and CPW-Based Coupling Structure}, 
  year={2019},
  volume={7},
  number={},
  pages={35089-35096},
  doi={10.1109/ACCESS.2019.2903147}}

@ARTICLE{Wei2020_BPF_SSPP_IEEE_plasma_sci,
  author={Wei, Yiwen and Wu, Yongle and Wang, Weimin and Pan, Leidan and Yang, Yuhao and Liu, Yuanan},
  journal={IEEE Transactions on Plasma Science}, 
  title={Double-Sided Spoof Surface Plasmon Polaritons- Line Bandpass Filter With Excellent Dual-Band Filtering and Wide Upper Band Suppressions}, 
  year={2020},
  volume={48},
  number={12},
  pages={4134-4143},
  doi={10.1109/TPS.2020.3035589}}

@ARTICLE{Liu-Xu2022_BPF_SSPP_80GHz_TMTT,
  author={Liu, Yiqun and Xu, Kai-Da and Li, Jianxing and Guo, Ying-Jiang and Zhang, Anxue and Chen, Qiang},
  journal={IEEE Transactions on Microwave Theory and Techniques}, 
  title={Millimeter-Wave E-Plane Waveguide Bandpass Filters Based on Spoof Surface Plasmon Polaritons}, 
  year={2022},
  volume={70},
  number={10},
  pages={4399-4409},
  doi={10.1109/TMTT.2022.3197593}}

@ARTICLE{Feng-Xu2024_BPF_SSPP_30GHz,
  author={Feng, Yinian and Xu, Kai-Da and Niu, Zhongqian and Zhang, Bo and Liu, Lujie and Fan, Yong},
  journal={IEEE Transactions on Microwave Theory and Techniques}, 
  title={Ka-Band Waveguide Bandpass Filters Using Double-Layer Grounded Spoof Surface Plasmon Polaritons}, 
  year={2024},
  volume={},
  number={},
  pages={1-11},
  doi={10.1109/TMTT.2024.3373504}}

@ARTICLE{THz_Communication_challanges_IEEE_TTST_2021,
  author={Song, Ho-Jin and Lee, Namyoon},
  journal={IEEE Transactions on Terahertz Science and Technology}, 
  title={Terahertz Communications: Challenges in the Next Decade}, 
  year={2022},
  volume={12},
  number={2},
  pages={105-117},
  doi={10.1109/TTHZ.2021.3128677}}

@INPROCEEDINGS{Haghighat_PACRIM_2024_6G_BPF,
  author={Haghighat, Mohsen and Darcie, Thomas and Smith, Levi},
  booktitle={2024 IEEE Pacific Rim Conference on Communications, Computers and Signal Processing (PACRIM)}, 
  title={A THz Spoof Surface Plasmon Polariton Band Pass Filter for 6G Communication Applications}, 
  year={2024},
  volume={},
  number={},
  pages={1-4},
  doi={10.1109/PACRIM61180.2024.10690197}}

\end{document}